\documentclass{llncs}

\usepackage{latexsym,bm}
\usepackage{graphicx}
\usepackage{subfigure}
\usepackage[noend]{algorithmic}
\usepackage{algorithm}
\usepackage{listings}
\usepackage{amsmath}

\floatname{algorithm}{Algorithm}

\newcommand{\tab}{\hspace*{1em}}

\begin{document}
\title{Scalable XSLT Evaluation\thanks{This work is supported in part
by the National Hi-Tech Research and Development Program of China
under Grant No.~2002AA116020 and by the National Natural Science Foundation
of China under Grant No.~60228006.}}
\author{Zhimao Guo \and Min Li \and Xiaoling Wang \and Aoying Zhou}
\institute{Dept.~of Computer Science and Engineering, Fudan University, China\\
\email{\{zmguo,leemin,wxling,ayzhou\}@fudan.edu.cn}}
\maketitle


\begin{abstract}
XSLT is an increasingly popular language for processing XML data.
It is widely supported by application platform software.
However, little optimization effort has been made inside the current XSLT processing engines.
Evaluating a very simple XSLT program on a large XML document with a simple schema may result
in extensive usage of memory.
In this paper, we present a novel notion of \emph{Streaming Processing Model} (\emph{SPM})
to evaluate a subset of XSLT programs on XML documents, especially large ones.
With SPM, an XSLT processor can transform an XML source document to other formats without
extra memory buffers required. Therefore, our approach can not only tackle
large source documents, but also produce large results.
We demonstrate with a performance study the advantages of the SPM approach.
Experimental results clearly confirm that SPM improves XSLT evaluation typically
2 to 10 times better than the existing approaches.
Moreover, the SPM approach also features high scalability.
\end{abstract}

\section{Introduction}
As XML has continued to gain popularity as a standard for
information representation and exchange, tools to process
XML are increasingly supported by common application platforms.
Many of them implement XSLT\cite{w3c:xslt}.
As a transformation language, XSLT has proven to be very popular with developers
and is often implemented as a stand-alone tool.
Unlike XQuery\cite{w3c:xquery}, XSLT was not
designed as a full-functional query language. Nevertheless, XSLT can easily be
used for query-like transformations.

When processing XML documents, most prevalent XSLT processors
try to keep the entire data structure of DOM or DOM-like models in main memory.
The size of DOM in the memory can be an order of manitude larger than
that of the original XML file.
Therefore, when dealing with XML files, if their size is comparable to or even larger than
the main memory, XSLT processors will thrash due to excessive use
of virtual memory. In this case, their efficiency degrades drastically.
These XSLT processing engines make little optimization effort.
As observed in our experiments,
even with respect to a rather simple XSLT program, the evaluation
process will be incredibly slow as long as the source XML document
is larger than the main memory.

In this paper, we present a novel notion of \emph{Streaming Processing Model} (\emph{SPM})
to evaluate a subset of XSLT programs on XML documents, especially large ones.
In the SPM model, given a DTD $\mathcal{D}$, an XSLT program $\mathcal{L}$ is converted to many handlers
for SAX-like events. When an XML file is read in, events will be fired.
At first, the result document is empty.
With the advancement of the scanning process, the output fragments generated by handlers of the events
will be appended to the result document continually.
Users obtained the result as soon as the scanning process was finished.
With our SPM model, an XSLT engine can transform an XML source document to other formats without
extra memory buffers required. Therefore, our approach can not only tackle
large source documents, but also produce large results.

\textbf{Contributions.}
In this paper, we made several contributions.
First, we proposed the SPM model to evaluate a subset of XSLTs on XML data.
With this model, XSLT can be evaluated without extra buffers required.
The content of the result document can be delivered continuously before all the source
data has been processed.

Second, we introduced \emph{transformation trees}, which incorporate the schema information
of the source document and the XSLT transformation.
We also devised algorithms to build a transformation tree and then convert it
into an SPM model.

Third, with a performance study, we demonstrated the advantages of the SPM approach.
Experimental results clearly confirmed that it improves XSLT evaluation typically 2
to 10 times faster than the current approaches, and also features high scalability.

\section{From DTD and XSLT to Transformation Tree}
In this section,
first we give the definition of \emph{simple DTDs} and \emph{XSLT$_{core}$},
then introduce \emph{transformation trees}.

DTD is well-known now.
For simplicity, we consider a set of simple DTDs in this paper.
A \emph{simple DTD} $\mathcal{D}$ contains no
IDs or IDREFs, and is acyclic.
Further, each element type has only one parent element type.
That is to say, in the DTD $\mathcal{D}$, if the element type $C$ is a child of the element type $A$,
then it cannot be a child of the element type $B$ at the same time.
Given a DTD, the last point can be easily satisfied by renaming some element names.
These assumptions guarantee that $\mathcal{D}$ can be represented as a DTD tree.

XSLT is a language for transforming XML documents into other formats.
In this paper, we do not consider the complete XSLT.
We define a restricted subset of XSLTs that we term \emph{XSLT$_{core}$},
which is described as following.

\textbf{XSLT$\boldsymbol{_{core}}$}.
An XSLT$_{core}$ program $\mathcal{L}$ is a set of template rules ${r_i}$, each of which is a
pair ($pattern(r_i)$, $template(r_i)$), where $pattern(r_i)$ is the match pattern
of $r_i$, $template(r_i)$ is the \emph{output template} of $r_i$ to form part of the result.
The output template is explained below.

\textbf{Output Template}.
An output template is a sequence $(o_1,o_2,\ldots,o_n)$ of two kinds of constructs: one is constant
strings, the other is apply-templates. That is to say, $o_i$ is either a constant string
or an apply-template.
We also assume that adjacent constant strings have been merged together into one longer string.
For each $i$, $o_i$ and $o_{i+1}$ cannot be constant strings simultaneously.

Combined with the definition of simple DTDs,
our definition of XSLT$_{core}$ further guarantees
that each template rule or apply-template can only match or select a single node or a set of
sibling nodes in the source document.
As said above, simple DTDs can be represented as DTD trees.
In a DTD tree, each node represents an element type, while each edge indicates
the parent/child relationship between two element types.
Given a simple DTD $\mathcal{D}$ and an XSLT$_{core}$ program $\mathcal{L}$,
$\mathcal{L}$ is evaluated on any instance $\mathcal{I}$ of $\mathcal{D}$.
Each template rule in $\mathcal{L}$ will be matched against a single node in the DTD tree of $\mathcal{D}$;
and each apply-template in template rules will select a single node of the DTD tree.

\begin{figure}[!htb]
\renewcommand{\baselinestretch}{1.0}
\centering
\lstset{basicstyle=\tiny,xleftmargin=0em,frame=single}
\begin{lstlisting}
<xsl:template match="/">
  <html><head><title>Books Information</title></head>
    <body><table>
      <xsl:apply-templates select="publication/book"/>
    </table></body>
  </html>
</xsl:template>
<xsl:template match="book">
  <tr><td><xsl:apply-templates select="title"/></td>
    <td><table>
        <xsl:apply-templates select="author"/>
    </table></td>
  </tr>
</xsl:template>
<xsl:template match="author">
  <tr><td><xsl:apply-templates select="name"/></td></tr>
</xsl:template>
<xsl:template match="title">
  <xsl:value-of select="."/>
</xsl:template>
<xsl:template match="name">
  <xsl:value-of select="."/>
</xsl:template>
\end{lstlisting}
\caption{An example of XSLT.}\label{fig:xslt}
\end{figure}
\begin{figure}[!htb]
\renewcommand{\baselinestretch}{1.0}
\centering
\lstset{basicstyle=\tiny,xleftmargin=0em,frame=single}
\begin{lstlisting}
<?xml version="1.0" encoding="UTF-8"?>
<publication>
  <book><title>A Complete Guide to DB2 Universal Database</title>
    <isbn>1-55860-482-0</isbn>
    <author><name>Don Chamberlin</name></author>
  </book>
</publication>
\end{lstlisting}
\caption{An XML document describing books information.}\label{fig:booksxml}
\end{figure}
\begin{figure}[!htb]
\renewcommand{\baselinestretch}{1.0}
\centering
\lstset{basicstyle=\tiny,xleftmargin=0em,frame=single}
\begin{lstlisting}
<!ELEMENT publication (book*)>
<!ELEMENT book (title,isbn,author*)>
<!ELEMENT title (#PCDATA)>
<!ELEMENT isbn (#PCDATA)>
<!ELEMENT author (name)>
<!ELEMENT name (#PCDATA)>
\end{lstlisting}
\caption{The DTD of the XML document in Fig.~\ref{fig:booksxml}.}\label{fig:booksdtd}
\end{figure}

\begin{figure}[!htb]
  \begin{minipage}[t]{0.5\linewidth}
    \centering
    \includegraphics[scale=0.4]{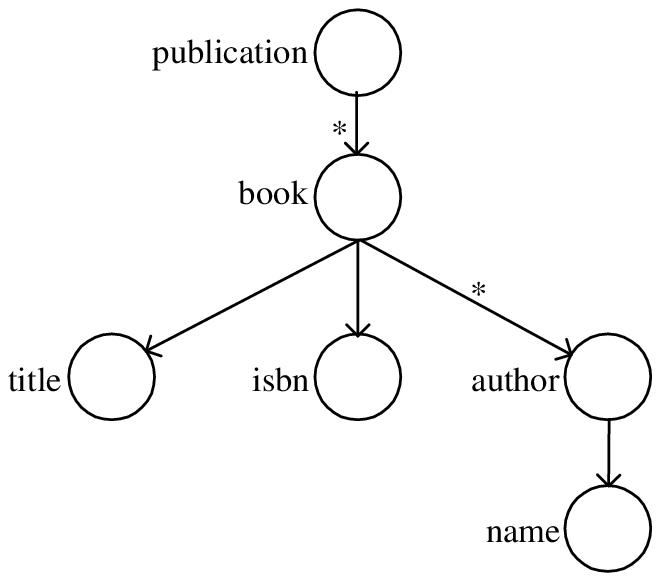}
    \caption{A DTD tree.}\label{fig:booksdtdtree}
  \end{minipage}%
  \begin{minipage}[t]{0.5\linewidth}
    \centering
    \includegraphics[scale=0.4]{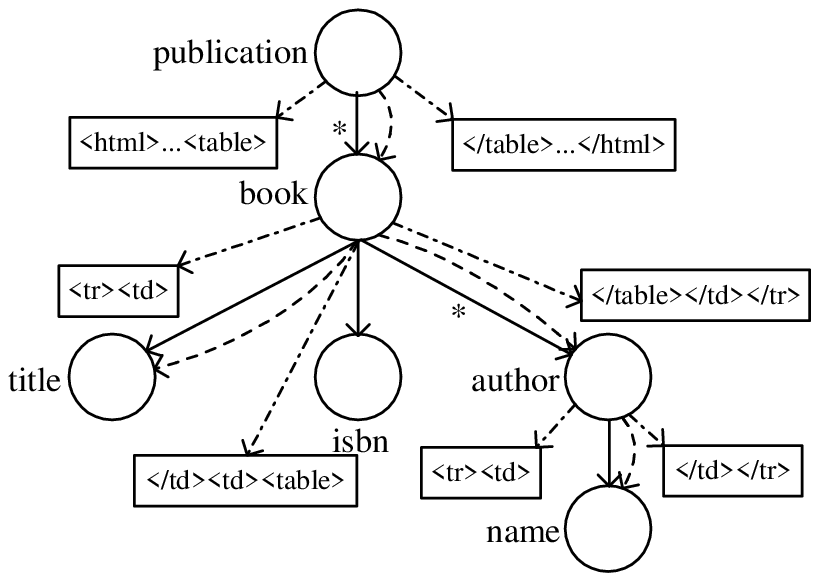}
    \caption{An transformation tree.}\label{fig:bookstt}
  \end{minipage}
\end{figure}

Fig.~\ref{fig:xslt} presents an XSLT$_{core}$ example.
Fig.~\ref{fig:booksxml} and~\ref{fig:booksdtd} shows an XML document and its DTD respectively,
while the DTD tree is illustrated in Fig.~\ref{fig:booksdtdtree}.
They will be used throughout this paper as a running example.
In order to integrate DTDs with XSLT$_{core}$ programs, we introduce a new data structure,
called \emph{transformation tree}, which is an extended DTD tree.
An example of transformation tree is depicted in Fig.~\ref{fig:bookstt},
which combines the information from the DTD in Fig.~\ref{fig:booksdtd} and the XSLT in Fig.~\ref{fig:xslt}.
In a transformation tree, a circle indicates an element type, while a rectangle indicates a constant
string from the XSLT program, and a dashed line represents an invocation
of certain apply-template.

Given the schema of the XML source document, denoted by its DTD in this paper, and an XSLT$_{core}$ program,
we devise algorithms to generate the corresponding transformation tree, which takes
both the schema information and the XSLT$_{core}$ into consideration. Next we
define \emph{transformation tree}, and present some notations.

\textbf{Transformation Tree}.
Given a simple DTD $\mathcal{D}$ and an XSLT$_{core}$ program $\mathcal{L}$,
the corresponding \emph{transformation tree} $\mathcal{T}$
is a rooted ordered tree.
It consists of two kinds of nodes: element nodes and constant string nodes.
Element nodes are indicated by circles, among which there is a distinguished root node,
while rectangles indicate constant string nodes.
There are three kinds of edges in $\mathcal{T}$. The first kind indicates
the parent/child relationship between two element types,
and is illustrated by solid lines.
As illustrated by dot-dashed lines,
the second kind of edges connects element nodes and constant string nodes.
The third kind represents the calling/being-called relationship,
and is illustrated by dashed lines in the diagram.
The first kind of edges can be derived from the DTD, while the XSLT program
introduces the latter two.

An XSLT program $\mathcal{L}$ is a set of template rules $r_i$, which consists of
two parts, match pattern $pattern(r_i)$ and output template $template(r_i)$.
Under the restrictions imposed on XSLT$_{core}$,
the $pattern(r_i)$ will be matched against a single node $n_i$ in the DTD tree.
We call $n_i$ the \emph{matched node} of $r_i$, denoted by $mnode(r_i)$.
As discussed above, $template(r_i)$ is a sequence $(o_{i1},o_{i2},\ldots,o_{it})$
of constant strings and apply-templates.
If $o_{ij} \: (1 \leq j \leq t)$ is an apply-template, then it will also select a single node
$n_{ij}$ in the DTD tree, denoted by $selnode(o_{ij})$.
We call $n_{ij}$ the \emph{selected node} of the apply-template $o_{ij}$.

\section{Streaming Processing of XSLT}
In this section, we describe the streaming processing model of XSLT evaluation
adopted in our work, and present the algorithms for converting a transformation tree
to a streaming processing model.

\begin{figure}[!htb]
  \centering
  \subfigure[]{
    \label{fig:xslt1}
    \includegraphics[scale=0.5]{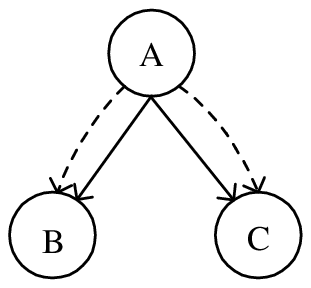}}
  \quad
  \subfigure[]{
    \label{fig:xslt2}
    \includegraphics[scale=0.5]{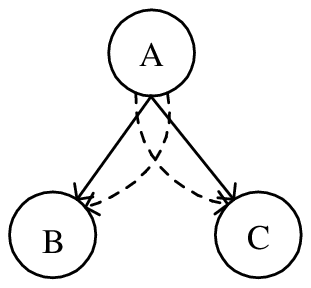}}
  \quad
  \subfigure[]{
    \label{fig:xslt3}
    \includegraphics[scale=0.5]{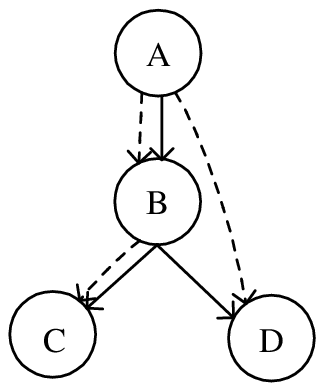}}
  \quad
  \subfigure[]{
    \label{fig:xslt4}
    \includegraphics[scale=0.5]{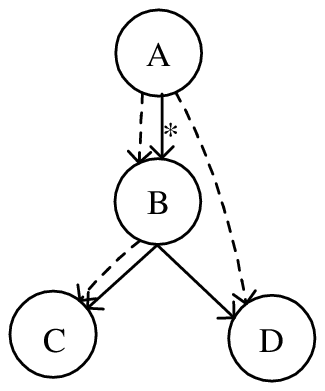}}
  \caption{Streamable vs.~unstreamable XSLTs.}\label{fig:pncase}
\end{figure}

Determining whether an XSLT program can be processed in a streaming fashion or not
is not easy. There are many subtle issues in the decision.
For brevity, before we give the strict definition of a \emph{streamable XSLT},
an XSLT program is said to be streamable, if it can be evaluated without extra buffers required.
In order to have a more clear understanding of this problem,
we first give several positive and negative cases, shown in Fig.~\ref{fig:pncase},
and present an enlightening discussion on them.

Let $\mathcal{D}_i$ and $\mathcal{L}_i$ denote the corresponding DTD and XSLT program
of the transformation tree$_i$, for $1 < i < 4$.
$\mathcal{D}_1$ is the same as $\mathcal{D}_2$, as shown below.
\lstset{basicstyle=\tiny,xleftmargin=0em,frame=single}
\begin{lstlisting}
<!ELEMENT A (B, C)>
<!ELEMENT B (#PCDATA)>
<!ELEMENT C (#PCDATA)>
\end{lstlisting}
\lstset{basicstyle=\normalsize,xleftmargin=0em,frame=single}
The root node of type $A$ has two children, one being of type $B$, and the other of type $C$.
$\mathcal{L}_1$ is rather simple.
It has only one non-trivial template rule $r$, the matched node of which is the root node.
Aside from constant strings\footnote{They are not shown in Fig.~\ref{fig:pncase}.},
$r$'s output template involves two apply-templates.
The first apply-template will select the child node of type B while the second
will select the child node of type C.
Comparing Fig.~\ref{fig:xslt2} with Fig.~\ref{fig:xslt1}, we see that
the order of two apply-templates is changed.
When transforming an XML document conforming to $\mathcal{D}_1$,
$\mathcal{L}_1$ can be evaluated without buffers required, while $\mathcal{L}_2$ cannot.
The latter needs extra buffers.
The reason for this is that the order $(B,C)$ in the XML streaming is opposite to
the order $(C,B)$ in the result tree of $\mathcal{L}_2$.
If no buffer can be used,
when the $B$ node comes, $\mathcal{L}_1$ will output something, while $\mathcal{L}_2$ will do nothing;
when the $C$ node comes, both $\mathcal{L}_1$ and $\mathcal{L}_2$ will output the \texttt{PCDATA} value of this node.
After that, $\mathcal{L}_1$ is well done.
However, $\mathcal{L}_2$ still has to access the node $B$, which has flowed away long before.
Therefore, $\mathcal{L}_2$ is not a streamable one with respect to $\mathcal{D}_2$.

$\mathcal{L}_3$ and $\mathcal{L}_4$ indicated by Fig.~\ref{fig:xslt3} and \ref{fig:xslt4}
is a bit more complicated than the former pair.
In fact, $\mathcal{L}_3$ is the same as $\mathcal{L}_4$,
but they are applied to XML documents conforming to different DTDs.
There is only a minor difference between $\mathcal{D}_3$ and $\mathcal{D}_4$.
In $\mathcal{D}_3$, $A$ has a child $B$, which appears only once;
while in $\mathcal{D}_4$, the edge from $A$ to $B$ is labelled by a `$*$', which indicates
that in the source document, the node of type $A$ can have zero or more children of type $B$.
In the case of Fig.~\ref{fig:xslt3}, part of the result document would be something like
``\texttt{<C>string$_1$</C> <D>string$_2$</D>}'',
which is the same as the internal order of the source document.
However, in the case of Fig.~\ref{fig:xslt4}, part of the result would be
``\texttt{<C>string$_1$</C>} \texttt{...} \texttt{<C>string$_2$</C>} \texttt{<D>string$_3$</D>}
\texttt{...} \texttt{<D>string$_4$</D>}''.
The order of these nodes appearing in the source document is ``\texttt{<C/>} \texttt{<D/>}
\texttt{...} \texttt{<C/>} \texttt{<D/>}'',
i.e., $C$ nodes and $D$ nodes occur alternately.
We cannot obtain the sequence $(C,C,\ldots,D,D)$ from the sequence $(C,D,C,D,\ldots,C,D)$,
if only a single pass of the original sequence is allowed and no memory buffers can be made use of
during the transformation process.

Next we present the definition of \emph{a streamable XSLT}.

\textbf{Streamable XSLT}.
Given a simple DTD $\mathcal{D}$, an XSLT$_{core}$ program $\mathcal{L}$ is said to be streamable
if it satisfies that,
for any XML instance $\mathcal{I}$ of $\mathcal{D}$, $\mathcal{L}$ can always be evaluated
successfully on $\mathcal{I}$ to produce
the correct result document with no memory buffers required.

Now it is the appropriate time to introduce the definition
of our streaming processing model.
When an XML document is scanned from its beginning to its end, a series of
events will be emitted. These events can be classified into
two categories, \emph{element-start} and \emph{element-end} events.
For simplicity, we do not consider attributes of elements in this work,
or assume that we have replaced them by sub-elements and modified the DTD correspondingly.
Next we describe the streaming processing model, shortened as SPM.

\textbf{Streaming Processing Model (SPM)}.
For each event $e$, no matter being element-start or element-end one,
there is an \emph{output fragment} attached to $e$.

To each element-start event, a constant string $cstr_s$, which may be \emph{empty}, is attached.
For a non-leaf node, there is also a constant string $cstr_e$ attached to its element-end event.
These strings constitute the output fragment of the event $e$.

Different from the above cases, a tri-tuple $(cstr_{e1},pcdata,cstr_{e2})$ is
attached to the element-end event of each leaf node,
here $pcdata$ field is either \emph{null} or the \texttt{PCDATA} value of the leaf node.
In this case, the concatenation of $cstr_{e1}$, $pcdata$ and $cstr_{e2}$ forms
the output fragment of the corresponding event.

When an XSLT program $\mathcal{L}$ is evaluated on a source document, a sequence of events
will be fired. The concatenation of their output fragments can be proven to be
the result document.
With the advancement of scanning process, output fragments can be appended to the result.
Hence, the result can be delivered continuously.
Finally, when the scanning process finished, the result document was obtained.

As can be seen from Fig.~\ref{fig:pncase}, given a DTD $\mathcal{D}$, an XSLT$_{core}$ program $\mathcal{L}$
may be streamable or not w.r.t.~$\mathcal{D}$.
In other words, it can be converted to a streaming
processing model or not. Thus, we have to consider what XSLT programs
can be converted and how they can be converted.

We devise the \textsc{Convert} procedure which tries to convert an XSLT program
into a streaming processing model. If it reports a failure,
this means that the XSLT program cannot be successfully converted to a streaming processing model.
In this paper, we will not separately present the algorithm for testing the streamability of
an XSLT program, and we deem that checking by theoretical proofs is feasible.
In fact, that is a major part of our future work.

The parameters of the algorithm \textsc{Convert} are an DTD $\mathcal{D}$ and an XSLT program $\mathcal{L}$,
and it returns a streaming processing model $\mathcal{M}$.
It first builds a transformation tree $\mathcal{T}$ from $\mathcal{D}$ and $\mathcal{L}$,
then builds the corresponding SPM model $\mathcal{M}$ through the procedure \textsc{BuildSPM}, which
is illustrated in Alg.~\ref{alg:buildspm}.
Before calling the procedure \textsc{BuildSPM},
it initializes output fragments of the element-start and element-end events of each node
$n$ in the transformation tree to empty strings.

\begin{figure}[!tb]
  \centering
  \subfigure[]{
    \label{fig:spm1}
    \includegraphics[scale=0.7]{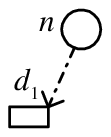}}
  \subfigure[]{
    \label{fig:spm2}
    \includegraphics[scale=0.7]{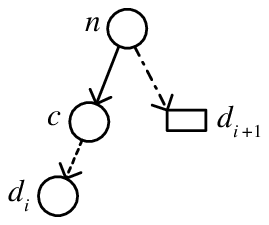}}
  \subfigure[]{
    \label{fig:spm3}
    \includegraphics[scale=0.7]{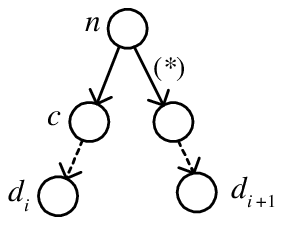}}
  \subfigure[]{
    \label{fig:spm4}
    \includegraphics[scale=0.7]{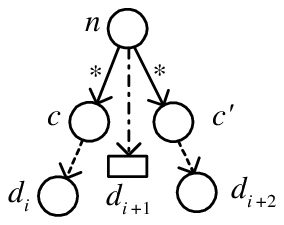}}
  \subfigure[]{
    \label{fig:spm5}
    \includegraphics[scale=0.7]{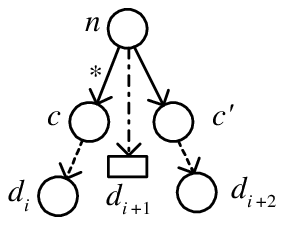}}
  \caption{Some concrete cases of the \textsc{BuildSPM} algorithm.}\label{fig:spm1to5}
\end{figure}

The input parameter of \textsc{BuildSPM} is a node $n$ of the transformation tree.
The first time the \textsc{BuildSPM} procedure is called by the procedure \textsc{Convert} rather than
by itself recursively, the root node of the transformation tree $\mathcal{T}$ is passed to $n$.
And during the entire recursive process, $\mathcal{M}$ is a global variable.
Let $start(n), end(n)$ denote the output fragments of the element-start and element-end events of the node $n$,
respectively.
$start(n)$ may be modified by \textsc{BuildSPM}($n_0$) or \textsc{BuildSPM}(n),
here $n_0$ is the parent node of $n$.
In \textsc{BuildSPM}($n_0$), something may be pre-appended to $start(n)$;
while in \textsc{BuildSPM}($n$), more may be post-appended to it.
In the pseudo-code, the operator `$+$' implies concatenation of strings,
e.g.\ $end(c) \leftarrow end(c) + d_{i+1}$, or addition of integer variables.
The analysis on $end(n)$ is similar, thus we do not detail it further.
We do not explain the algorithm \textsc{BuildSPM} line by line.
Instead, we present several illustrations to facilitate understanding.
Concrete cases related to lines 5--7, 11--12, 13, 17--18 and 19--20 of Alg.~\ref{alg:buildspm}
are shown in Fig.~\ref{fig:spm1}, \ref{fig:spm2}, \ref{fig:spm3}, \ref{fig:spm4} and \ref{fig:spm5}, respectively.

\begin{algorithm}[!htb]
\small
\caption{\textsc{BuildSPM}}\label{alg:buildspm}
\begin{algorithmic}[1]
\REQUIRE node $n$
\STATE \textbf{if} ($n$ is a leaf node) \textbf{then}
\STATE \tab $end(n) \leftarrow pcdata + end(n)$; \textbf{return}
\STATE \COMMENT{$c_j\: (1 \le j \le p)$ to denote children of $n$ derived from DTD.}
\STATE \COMMENT{$d_i\: (1 \le i \le q)$ to denote children of $n$ introduced by XSLT.}
\STATE \textbf{if} $d_1$ is a constant string node
\STATE \tab $start(n) \leftarrow start(n) + d_1$; $i \leftarrow 2$
\STATE \textbf{else} $i \leftarrow 1$
\STATE \textbf{while} $i \le q$
\STATE \tab $c \leftarrow$ the child of $n$ along the path from $n$ to $d_i$
\STATE \tab \textbf{if} the edge $(n,c)$ is not labelled by `$*$'
\STATE \tab \tab \textbf{if} $d_{i+1}$ is a constant string node
\STATE \tab \tab \tab $end(c) \leftarrow end(c) + d_{i+1}$; $ i \leftarrow i + 2$
\STATE \tab \tab \textbf{else} $ i \leftarrow i + 1 $
\STATE \tab \textbf{else}
\STATE \tab \tab \textbf{if} $d_{i+1}$ is a constant string node
\STATE \tab \tab \tab $c' \leftarrow$ the child of $n$ along the path from $n$ to $d_{i+2}$
\STATE \tab \tab \tab \textbf{if} $(n,c')$ is labelled by `$*$'
\STATE \tab \tab \tab \tab Err(``not streamable!'')
\STATE \tab \tab \tab \textbf{else}
\STATE \tab \tab \tab \tab $start(c') \leftarrow d_{i+1} + start(c')$; $i \leftarrow i + 2$
\STATE \tab \tab \textbf{else} $i \leftarrow i + 1$
\STATE \textbf{for} each $d_i$ of element node type
\STATE \tab BuildSPM($d_i$)
\end{algorithmic}
\end{algorithm}

As a complete example, the transformation tree in Fig.~\ref{fig:bookstt} will
be converted to the following SPM model.
\lstset{basicstyle=\tiny,xleftmargin=0em,frame=single}
\begin{lstlisting}
start(publication)="<html>...<table>"
end(publication)="</table>...</html>"
start(book)="<tr><td>"
end(book)="</table></td></tr>"
end(title)=concat(PCDATA value of title, "</td><td><table>")
begin(author)="<tr><td>"
end(name)=concat(PCDATA value of name, "</td><tr>")
\end{lstlisting}
\lstset{basicstyle=\normalsize,xleftmargin=0em,frame=single}
Note that trivial output fragments are not shown here,
such as \emph{start(title)=``"}, etc.

As can be understood from the above SPM model,
an output fragment is attached to each element-start and element-end event.
During the scanning process of the source document, many output fragments will be generated.
Regarding to streamable XSLT programs, no extra memory buffer is required
when evaluating them on source documents.
Before all the source data has arrived, the head part of the result can be delivered.
Even a large XML data instance can be processed with only a single pass.

\section{Experimental Results}
In this section, we report the experimental results.
We examined the performance of our SPM approach on XML data of different sizes,
and compared it to several publicly available XSLT processing engines.
We evaluated an XSLT transformation $\mathcal{L}_0$ on the DBLP XML documents.
Below we present the result for $\mathcal{L}_0$ that transforms a DBLP
XML document into another schema: for each conference paper, an HTML table row
is generated,
listing all authors of this paper in a nested table,
followed by the title of the paper.
$\mathcal{L}_0$ is much like a query, extracting part of data,
then tagging them in a different way from the source document.
Due to space constraint, $\mathcal{L}_0$ is not shown here.

Our experiments were carried out on a PC with an Intel \emph{PIV} 1.8GHz processor
and 256MB of RAM, running MS Windows 2000 Server.
MS command line XSLT tool v1.1 was used.
And Sun JDK v1.4 was used as the Java runtime environment.
Xalan-j v2.5.1 and saxon v6.5.3 were also used for the experiments.
Our algorithms related to the SPM model are all implemented in JDK 1.4.
We used varying sizes of XML source
documents from the well-known DBLP database.

\begin{figure}[!htb]
  \centering
  \subfigure[]{
    \label{fig:meddata}
    \includegraphics[scale=0.4]{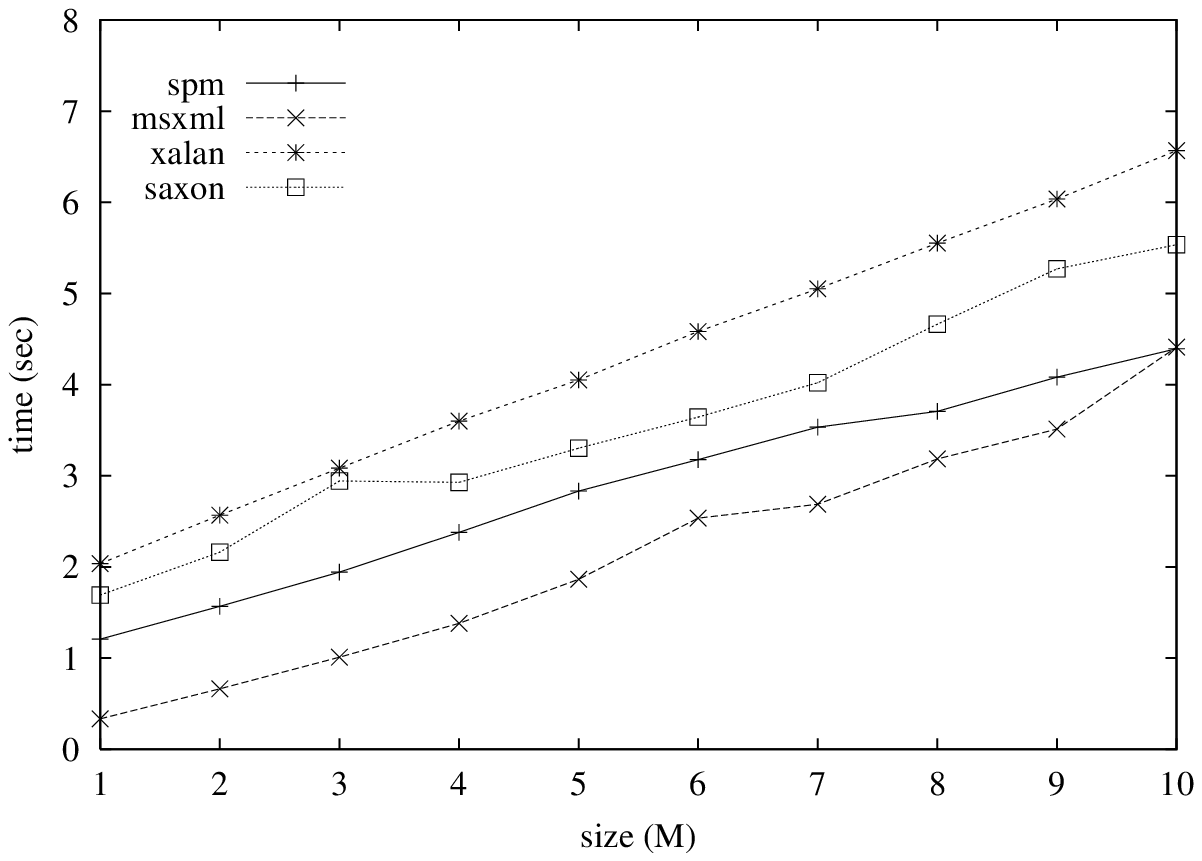}}
  \subfigure[]{
    \label{fig:largedata}
    \includegraphics[scale=0.4]{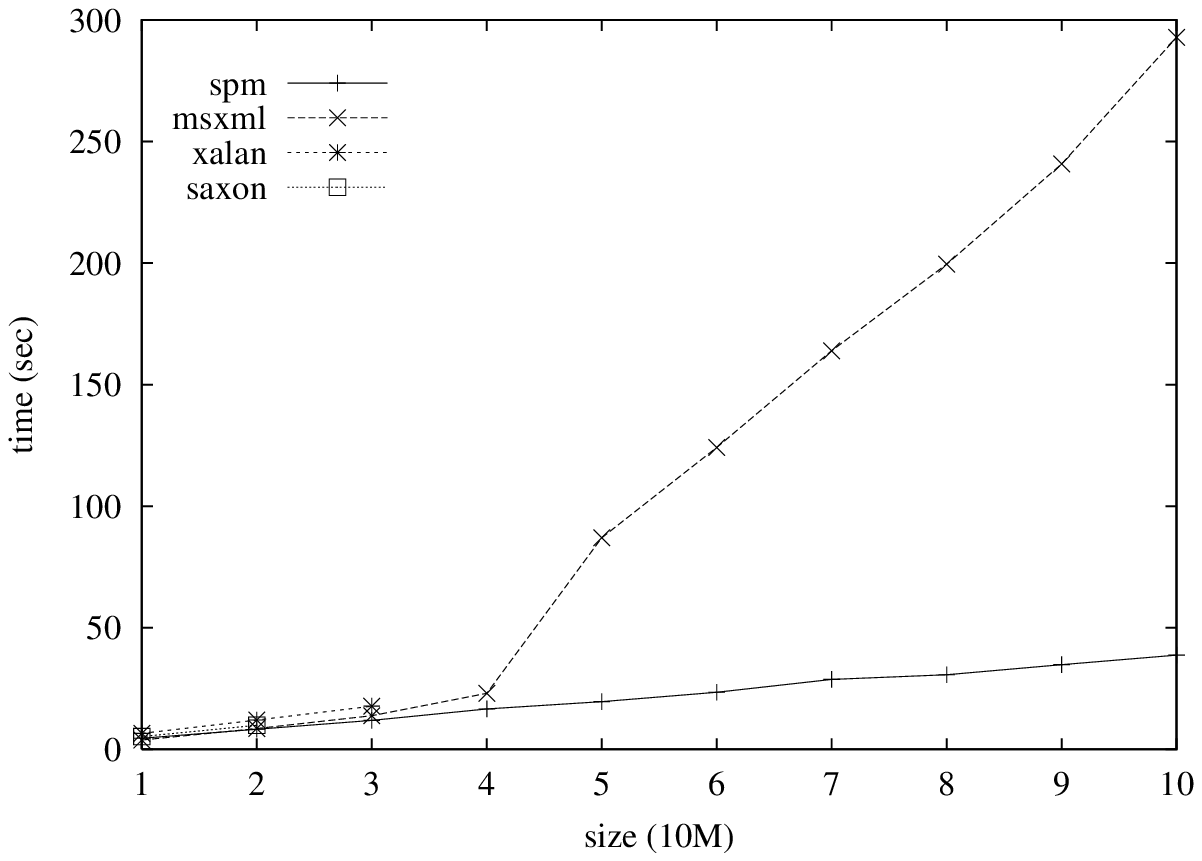}}
  \caption{Evaluation time vs.~size of DBLP source documents.}\label{fig:expresult}
\end{figure}

We conducted two groups of experiments.
The first group is a comparative study.
In it, we compare the performance of Xalan-j, Saxon, MS XML and SPM
on DBLP XML documents, whose sizes range from 1\emph{M} to 10\emph{M}.
The evaluation time measures both the time to parse XSLTs (and DTDs)
as well as the time to carry out the transformation.
As can be seen from Fig.~\ref{fig:meddata}, in this scale of source document sizes,
MS XSLT tool performs best, followed by our SPM approach, then Saxon and Xalan-j.
With the increase of document sizes, the gap lying between MS XSLT tool and
SPM becomes less obvious.
It is expected that the SPM approach will perform better than MS XSLT tool,
if the input document is larger than 10\emph{M}.
That is confirmed by the other group of experiments.

The second group is a scalability test, in which we investigate
the scalability of the SPM approach in comparison with MS XML, Saxon and Xalan-j
using large XML documents, whose sizes range from 10\emph{M} to 100\emph{M}.
Fig.~\ref{fig:largedata} depicts the corresponding results.
When the XML document is larger than 20\emph{M}, Saxon processor reports
\emph{out of the memory} exception; so does Xalan-j processor, when the document
is larger than 30\emph{M}.
Though MS XSLT tool does not throw exceptions,
it causes extensive usage of virtual memory, which goes beyond 580\emph{M} sometimes.
Out of the 10 repeated runs of the same configuration, it reports \emph{deficiency in virtual memory} once.
As to the SPM approach, however, the evaluation time grows approximately linearly with
the increase of the document size.
Fig.~\ref{fig:largedata} clearly demonstrates that SPM improves XSLT evaluation 2 to 10 times better than MS XML.
Therefore, high scalability of our SPM approach is confirmed.

\section{Related Work and Conclusion}
The problem of incorporating XSLT processing into database engines
is investigated in~\cite{moerkotte:xsl2db}.
The approach in~\cite{jain:translate} generates a single SQL
for an XSLT program and works for a large fragment of XSLT.
\cite{li:xslt} studies how to compose an XSLT transformation with an XML view.
A major distinction of our work from \cite{moerkotte:xsl2db,jain:translate,li:xslt} is that
their work is based on databases while we focus on processing XML documents directly.

Much of the previous work is devoted to evaluating XPath queries over streaming
XML\cite{olteanu:lookforward,barton:withbackward,altinel:xfilter,diao:yfilter,chan:xtrie,sf:xmltk,green:automata}.
However, to the best of our knowledge, our work is the first effort
to study streaming processing in the context of XSLT evaluation.

Our approach supports scalable XSLT evaluation.
The SPM model offers several novel features that make it
especially attractive for transforming large XML data.
Our experimental results have clearly demonstrated the benefits
of our approach.
They show that the SPM approach outperforms current XSLT processors
by an order of magnitude when dealing with large XML documents.


\begin{thebibliography}{10}

\bibitem{sf:xmltk}
{xmltk: An XML Toolkit for Lightweight XML Stream Processing}.
\newblock http://xmltk.sourceforge.net/.

\bibitem{altinel:xfilter}
M.~Altinel and M.~J. Franklin.
\newblock {Efficient Filtering of XML Documents for Selective Dissemination of
  Information}.
\newblock In {\em Proc.~of VLDB}, 2000.

\bibitem{barton:withbackward}
C.~Barton, P.~Charles, D.~Goyal, et~al.
\newblock {Streaming XPath Processing with Forward and Backward Axes}.
\newblock In {\em Proc.~of ICDE}, 2003.

\bibitem{chan:xtrie}
C.~Y. Chan, P.~Felber, M.~N. Garofalakis, and R.~Rastogi.
\newblock {Efficient Filtering of XML Documents with XPath Expressions}.
\newblock In {\em Proc.~of ICDE}, 2002.

\bibitem{diao:yfilter}
Y.~Diao, P.~Fischer, M.~J. Franklin, et~al.
\newblock {YFilter: Efficient and Scalable Filtering of XML Documents}.
\newblock In {\em Proc.~of ICDE}, 2002.

\bibitem{green:automata}
T.~J. Green, G.~Miklau, M.~Onizuka, and D.~Suciu.
\newblock {Processing XML Streams with Deterministic Automata}.
\newblock In {\em Proc.~of ICDT}, pages 173--189, 2003.

\bibitem{jain:translate}
S.~Jain, R.~Mahajan, and D.~Suciu.
\newblock {Translating XSLT Programs to Efficient SQL Queries}.
\newblock In {\em Proc.~of WWW}, pages 616--626, 2002.

\bibitem{li:xslt}
C.~Li, P.~Bohannon, H.~F. Korth, and P.~Narayan.
\newblock {Composing XSL Transformations with XML Publishing Views}.
\newblock In {\em Proc.~of SIGMOD}, pages 515--526, 2003.

\bibitem{moerkotte:xsl2db}
G.~Moerkotte.
\newblock {Incorporating XSL Processing into Database Engines}.
\newblock In {\em Proc.~of VLDB}, pages 107--118, 2002.

\bibitem{olteanu:lookforward}
D.~Olteanu, H.~Meuss, T.~Furche, et~al.
\newblock {XPath: Looking Forward}.
\newblock In {\em Proc.~of the EDBT Workshop on XML Data Management (XMLDM)},
  2002.

\bibitem{w3c:xquery}
{W3C}.
\newblock {XQuery 1.0: An XML Query Language}.
\newblock http://www.w3.org/TR/xquery/.

\bibitem{w3c:xslt}
{W3C}.
\newblock {XSL Transformations (XSLT) Version 1.0}.
\newblock http://www.w3.org/TR/xslt/.

\end{thebibliography}


\end{document}